\documentclass[twocolumn,showpacs,amsmath,amssymb,prc,superscriptaddress,floatfix]{revtex4}
\usepackage{graphicx}
\usepackage{amsfonts}
\usepackage{color}
\usepackage[T1]{fontenc}
\usepackage{lmodern}

\newcommand{\dd}{\mbox{$\textrm{d}$}}

\begin{document}
\renewcommand{\theenumi}{(\alph{enumi})}

\title{Determination of the $\boldsymbol{\eta ^3{\rm He}}$ threshold structure from the low energy $\boldsymbol{pd \to \eta ^3{\rm He}}$ reaction}

\date{\today}

\author{Ju-Jun Xie}~\email{xiejujun@impcas.ac.cn}%
\affiliation{Institute of Modern Physics, Chinese Academy of
Sciences, Lanzhou 730000, China}%
\affiliation{Departamento de F\'{\i}sica Te\'orica and IFIC, Centro Mixto
Universidad de Valencia-CSIC Institutos de Investigaci\'on de Paterna, Aptdo.
22085, 46071 Valencia, Spain}
\author{Wei-Hong Liang}~\email{liangwh@gxnu.edu.cn}%
\affiliation{Department of Physics, Guangxi Normal University,
Guilin 541004, China}
\author{Eulogio Oset}
\affiliation{Departamento de F\'{\i}sica Te\'orica and IFIC, Centro
Mixto Universidad de Valencia-CSIC Institutos de Investigaci\'on de
Paterna, Aptdo. 22085, 46071 Valencia, Spain}%
\affiliation{Institute of Modern Physics, Chinese Academy of Sciences,
Lanzhou 730000, China}%
\author{Pawel Moskal}
\affiliation{Institute of Physics, Jagiellonian University, ul.
Stanis{\l}awa {\L}ojasiewicza 11, 30-348 Krak\'ow, Poland}
\author{Magdalena Skurzok}%
\affiliation{Institute of Physics, Jagiellonian University, ul.
Stanis{\l}awa {\L}ojasiewicza 11, 30-348 Krak\'ow, Poland}
\author{Colin Wilkin} %
\affiliation{Physics and Astronomy Department, UCL, London WC1E 6BT,
UK}

\begin{abstract}

We analyze the data on cross sections and asymmetries for the $pd
(dp) \to \eta ^3{\rm He}$ reaction close to threshold and look for
bound states of the $\eta ^3 {\rm He}$ system. Rather than
parameterizing the scattering matrix, as is usually done, we develop
a framework in which the $\eta ^3 {\rm He}$ optical potential is the
key ingredient, and its strength, together with some production
parameters, are fitted to the available experimental data. The
relationship of the scattering matrix to the optical potential is
established using the Bethe-Salpeter equation and the $\eta ^3 {\rm
He}$ loop function incorporates the range of the interaction given
by the empirical $^3 {\rm He}$ density. We find a local Breit Wigner
form of the $\eta^3$He amplitude $T$ below threshold with a clear
peak in $|T|^2$, which corresponds to an $\eta^3 {\rm He}$ binding
of about 0.3~MeV and a width of about 3~MeV. By fitting the
potential we can also evaluate the $\eta ^3 {\rm He}$ scattering
length, including its sign, thus resolving the ambiguity in the
former analyses.
\end{abstract}

\pacs{21.85.+d, 
      14.40.Aq,    
      13.75.-n 
}

\maketitle

\section{Introduction}

The identification of $\eta$ bound states in nuclei has been a constant
thought for several
years~\cite{Wilkin:2016ajz,Bass:2015ova,Haider:2015fea,Kelkar:2015bta,Hirenzaki:2015eoa,Friedman:2013zfa},
starting from the early works of Refs.~\cite{Bhaleliu,Haiderliu,Liuhaider}.
More precise evaluations of the $\eta$-nucleus optical potential, with
special attention to two-nucleon $\eta$ absorption, indicated that, while
indeed the $\eta N$ interaction was strong enough to bind $\eta$ states, the
widths were always bigger than the binding~\cite{chiang}.

An important step forward was made possible by the advent of chiral unitary
theory to describe the meson-baryon
interaction~\cite{Kaiser,ramos,ollerulf,Nieves,Hyodo}. Within this
theoretical framework, the $\eta N$ interaction with coupled channels was
studied in Refs.~\cite{Kaiser,Inoue}, and the $N^*(1535)$ resonance appeared
dynamically generated in the scheme, albeit with unnatural subtraction
constants in the regularization of the loops. This deficiency is an
indication that other components are missing in the
approach~\cite{Jido,Sekihara}, and the problem was solved in
Ref.~\cite{Garzon} through the inclusion of the $\rho N$ and $\pi \Delta$
channels, in addition to the $\pi N, \eta N, K\Lambda, K \Sigma$ of the
original chiral unitary approach.

The $\eta$-nucleus interaction within the chiral unitary approach was studied
in Ref.~\cite{Inoueta}, where enough attraction was found to form bound
$\eta$-nucleus states. Detailed studies of the $\eta$ energies for different
nuclei were made in Ref.~\cite{GarciaRecio:2002cu}, where for medium and
light nuclei, bound states were found (see also Ref.~\cite{Cieply:2013sga},
where qualitatively similar conclusions were drawn), though with larger
widths than binding energies. For instance, for $^{12}\textrm{C}$, binding
$B_E = 9.7~\textrm{MeV}$ and width $\Gamma=35~\textrm{MeV}$ were found with
the preferred energy-dependent potential. Assuming that these two magnitudes
scale roughly with the mass number, for $^3\textrm{He}$ we could expect $B_E
\simeq 2.4~\textrm{MeV}$, $\Gamma\simeq 8.7~\textrm{MeV}$, though the width
could be somewhat smaller than this because the relative weight of two-body
$\eta$ absorption should be smaller in $^3\textrm{He}$ than in
$^{12}\textrm{C}$. On the other hand, some theoretical calculations for light
systems predict $B_E$ of around 1 MeV or less and $\Gamma = 15$~MeV for $\eta
^3{\rm He}$~\cite{Barnea:2015lia}. The fact that the widths are expected to
be much larger than the binding might be the reason why so far, we have no
conclusive evidence for any of these bound
states~\cite{Bilger:2002aw,Mersmann,Colin,Rausmann:2009dn,Urban:2009zzc,
Chrien:1988gn,Johnson:1993zy,Willis:1997ix,Sokol:1998ua,Budzanowski:2008fr,
Moskal:2010ee,Pheron:2012aj,Adlarson:2013xg,Fujioka:2015pla,Skurzok:2016fuv}.

The data on the $pd(dp) \to \eta~^3\rm{He}$ total cross section show a sharp
rise from threshold before becoming stable at an excess energy of about
$Q=1$~MeV, keeping this constant value up to about
10~MeV~\cite{Mersmann,Smyrski:2007nu}. These data have been analyzed before
in Refs.~\cite{Mersmann,Colin}. In Ref.~\cite{Mersmann} only an $s$-wave
amplitude for $\eta~^3\textrm{He}$ was considered, while in Ref.~\cite{Colin}
the $s$-wave and $p$-wave interference data were considered in order to
further constrain the $\eta~^3\textrm{He}$ amplitude. This analysis suggested
a pole with a binding energy of around 0.3~MeV and with a very small width.

In the present work, we describe an alternative method of analysis,
following the algorithms used in the chiral unitary approach. This
allows one to produce an $\eta~^3\textrm{He}$ amplitude that is
fully unitary and with proper analytical properties, without the
approximations, or assumptions, made in Refs.~\cite{Mersmann,Colin}.
Our approach does not pre-assume any particular form of the
amplitude but rather generates it from an $\eta~^3\textrm{He}$
potential, which is what is fitted to the data. The $T$-matrix then
arises from the solution of the Lippmann-Schwinger equation though,
for convenience, we use the Bethe-Salpeter equation (BSE), which
allows us to keep relativistic terms.

In our approach we can relate the parameters of the potential to the $\eta N$
scattering length and this provides a valuable constraint. The $\eta N$
scattering length in the chiral unitary approach is estimated to be $a_{\eta
N}=(-0.264-i0.245) ~\rm{fm}$ in Ref.~\cite{Inoueta}, and $a_{\eta
N}=(-0.20-i0.26) ~\rm{fm}$ in Ref.~\cite{Kaiser}. Other approaches include
the results of Ref.~\cite{Wycech} with $a_{\eta N}=(-0.87-i0.27) ~\rm{fm}$
and those of Ref.~\cite{Batinic:1995yu} with $a_{\eta N}=(-0.691-i0.174)
~\rm{fm}$ in one option and $a_{\eta N}=(-0.968-i0.281) ~\rm{fm}$ in another.
A different version by the same group yielded $a_{\eta N}=(-0.910 \pm 0.050 -
i(0.290 \pm 0.04)) ~\rm{fm}$~\cite{Batinic:1995kr}. An interesting result is
the constraint on ${\rm Im}(a_{\eta N})$ from the optical theorem and the
inelastic cross section of $\pi N \to \eta N$~\cite{Batinic:1995kr}, $|{\rm
Im}(a_{\eta N})| \geq (0.24 \pm 0.02) ~ {\rm fm}$.

Some parameters that we fit to the data can be related, at least
approximately, to $a_{\eta N}$, and this will be used as a consistency check
of the results. As we shall see later, the output of our calculations leads
to an $\eta~^3{\rm He}$ optical potential from which we deduce a value of
$a_{\eta N}$ that is basically consistent with experimental information. With
this optical potential we solved the BSE for the $\eta~^3\textrm{He}$ system
and found an $\eta~^3{\rm He}$ bound state with a binding energy of around
$0.3$~MeV and a width of around $3$~MeV, with reasonable uncertainties. We
can therefore claim that the data of the $pd(dp) \to \eta~^3\textrm{He}$
reaction close to threshold provide evidence of a very weakly bound
$\eta~^3\rm{He}$ state.

\section{Formalism}

\subsection{The $\boldsymbol{\eta~^3\rm{He}}$ interaction}

\begin{figure}[htbp]\centering
\includegraphics[width=1.0\columnwidth]{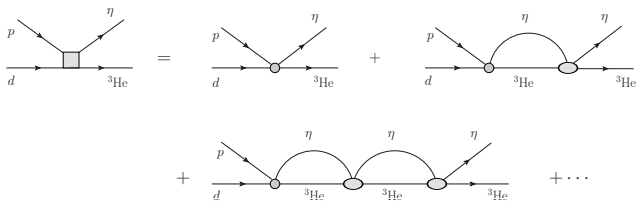}
\caption{The process $pd \to \eta~^3\textrm{He}$ considering
explicitly the $\eta~^3\textrm{He}$ rescattering. The square box in
the first diagram indicates the full transition amplitude, while the
circle in the second diagram stands for the bare transition
amplitude prior to the $\eta~^3\textrm{He}$ final state interaction.
It contains all diagrams that do not have $\eta ~ ^3{\rm He}$ as
intermediate state. The oval stands for the $\eta~^3\textrm{He}$
optical potential. \label{fig:fig1}}
\end{figure}

 Let us depict diagrammatically the $pd \to \eta~^3\textrm{He}$ process. This is done in
 Fig.~\ref{fig:fig1}.

\begin{figure}[htbp]\centering
\includegraphics[width=1.\columnwidth]{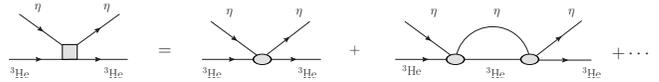}
\caption{Diagrammatic representation of the $\eta~^3\textrm{He}$
scattering matrix. \label{fig:fig2}}
\end{figure}

The $\eta~^3\textrm{He}$ scattering amplitude is given by the diagrams
depicted in Fig.~\ref{fig:fig2}, and formally by
\begin{equation}\label{eq:BSE}
T= V+VGT,
\end{equation}
where $V$ is the $\eta~^3\textrm{He}$ optical potential, which contains an
imaginary part to account for the inelastic channels $\eta~^3\textrm{He} \to
X$, where $X$ is mostly $\pi 3N$. It also includes the $3N$ intermediate
state arising mainly from $\eta$ two body absorption, $\eta NN \to
NN$~\cite{chiang,Inoueta,Wycech:2014wua}.

In many body theory it is known that at low densities the optical potential
is given by,
\begin{eqnarray}
\label{eq:vroptical}
V(\vec{r}) = 3t_{\eta N} \tilde{\rho}(\vec{r}),
\end{eqnarray}
where $t_{\eta N}$ is the forward $\eta N$ amplitude and
$\tilde{\rho}(\vec{r})$ is the $^3\rm{He}$ density normalized to unity.
Eq.~\eqref{eq:vroptical} is relatively accurate in many body physics, but
here we are more concerned with the fact that it provides the realistic range
of $\eta$-nucleus interaction, since the $\eta$ can interact with all the
nucleons in the nucleus distributed according to $\rho(\vec{r})$.

In momentum space the potential is given by
\begin{eqnarray}
\nonumber
V(\vec{p}_\eta,\vec{p'}_\eta) &=& 3t_{\eta N} \int \dd^3\vec{r}\,
\tilde{\rho}(\vec{r}) e^{i(\vec{p}_\eta - \vec{p'}_{\eta})\cdot
\vec{r}}\\ &=& 3t_{\eta N}F(\vec{p}_\eta - \vec{p'}_{\eta}),
\label{eq:opotential}
\end{eqnarray}
where $F(\vec{q})$ is the $^3\rm{He}$ form factor,
\begin{eqnarray}
F(\vec{q}) = \int \dd^3\vec{r}\, \tilde{\rho}(\vec{r}) e^{i\vec{q} \cdot
\vec{r}},
\end{eqnarray}
and $F(\vec{0}) = 1$. A good approximation to this form factor at small
momentum transfers is given by a Gaussian,
\begin{eqnarray}
F(\vec{q}) = e^{-\beta^2 |\vec{q}|^{2}},
\end{eqnarray}
where $\beta^2= \langle r^2\rangle/6$. This mean-square radius corresponds to
the distribution of the centers of the nucleons and, after correcting for the
nucleon size, this leads to an experimental value of $\beta^2 = 13.7 ~{\rm
GeV}^{-2}$~\cite{Sick:2014yha}.

Due to the form factor, the optical potential in Eq.~\eqref{eq:opotential}
contains all partial waves. There are other sources for the $p$-waves and
these will be treated later in an empirical way. After integrating over  the
angle between $\vec{p}^{~'}_\eta$ and $\vec{p}_\eta$, the $s$-wave projection
of the optical potential becomes
\begin{eqnarray}
V(\vec{p}_\eta,\vec{p}^{~'}_\eta) &=& 3 t_{\eta N}
\frac{1}{2}\int^1_{-1} \dd\cos\theta e^{-\beta^2
(|\vec{p}_\eta|^2 + |\vec{p'}_\eta|^2 - 2|\vec{p}_{\eta}|
|\vec{p'}_{\eta}|\cos\theta)} \nonumber\\
&&\hspace{-1.5cm}= 3 t_{\eta N} e^{-\beta^2 |\vec{p}_{\eta}|^2} e^{-\beta^2
|\vec{p'}_{\eta}|^2} \left[1 + \frac{1}{6}
(2\beta^2|\vec{p}_{\eta}||\vec{p'}_{\eta}|)^2 + ... \right]\!.
\end{eqnarray}
The term $2\beta^2|\vec{p}_{\eta}||\vec{p'}_{\eta}|/6$ is negligible
in the region where $e^{-\beta^2 |\vec{p}_{\eta}|^2} e^{-\beta^2
|\vec{p'}_{\eta}|^2}$ is sizeable and can be neglected and this
leads to a potential that is separable in the variables
$\vec{p}_\eta$ and $\vec{p}^{~'}_\eta$, which makes the solution of
Eq.~\eqref{eq:BSE} trivial. Keeping the relativistic factors of the
Bethe Salpeter equation, we can write~\cite{ramos}:
\begin{eqnarray}
 \label{eq:Tamplitude} T(\vec{p}_\eta,\vec{p'}_\eta) &=& \tilde{V} e^{-\beta^2
|\vec{p}_{\eta}|^2} e^{-\beta^2 |\vec{p'}_{\eta}|^2}  + \\
\nonumber && \hspace{-2.2cm}\int \frac{\dd^3\vec{q}}{(2\pi)^3}
\frac{M_{^3{\rm He}}}{2\omega_{\eta}(\vec{q}) E_{^3{\rm
He}}(\vec{q})} \frac{\tilde{V} e^{-\beta^2 |\vec{p}_{\eta}|^2}
e^{-\beta^2 |\vec{q}\hspace{0.5mm}|^2}}{\sqrt{s} -
\omega_{\eta}(\vec{q}) - E_{^3{\rm He}}(\vec{q}) + i\epsilon}
T(\vec{q},\vec{p'}_\eta),
\end{eqnarray}
with $\sqrt{s}$ being the invariant mass of the $\eta~^3{\rm He}$ system,
$\omega_{\eta}(\vec{q}) = \sqrt{m^2_{\eta} + |\vec{q}\hspace{0.5mm}|^2}$, and
$E_{^3{\rm He}} (\vec{q}) = \sqrt{M^2_{^3{\rm He}} +
|\vec{q}\hspace{0.5mm}|^2}$. We have here taken $\tilde{V}$ instead of
$3t_{\eta N}$ for more generality.

By expanding $T$ in Eq.~\eqref{eq:Tamplitude} as $V + VGV + VGVGV +
...$ we can see that all terms with $G$ contain $e^{-2\beta^2
|\vec{q}|^2}$ and that the factors $e^{-\beta^2 |\vec{p}_{\eta}|^2}
e^{-\beta^2 |\vec{p'}_{\eta}|^2}$ can be factorized outside the
integral. Hence, the $T$ matrix can be factorized in the same way as
$V$, and we have
\begin{eqnarray}
T(\vec{p}_\eta,\vec{p'}_\eta) = \tilde{T} e^{-\beta^2
|\vec{p}_{\eta}|^2} e^{-\beta^2 |\vec{p'}_{\eta}|^2}.
\end{eqnarray}
The Bethe Salpeter equation becomes then algebraic
\begin{eqnarray}
\tilde{T} = \tilde{V} + \tilde{V} G \tilde{T}, \label{eq:ttilde}
\end{eqnarray}
with
\begin{eqnarray}
&& G = \frac{M_{^3{\rm
He}}}{16\pi^3} \times \nonumber \\
&& \int \frac{\dd^3\vec{q}}{\omega_\eta(\vec{q})E_{^3{\rm
He}}(\vec{q})} \frac{e^{-2 \beta^2 |\vec{q}|^2}}{\sqrt{s} -
\omega_{\eta}(\vec{q}) - E_{^3{\rm He}}(\vec{q}) + i\epsilon}.
\label{eq:gfunction}
\end{eqnarray}

\begin{figure}[htbp]\centering
\includegraphics[width=1.0\columnwidth]{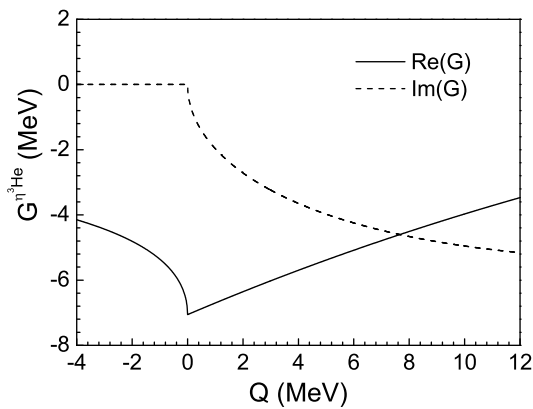}
\caption{Real (solid line) and imaginary (dashed line) parts of the $G$ of
Eq.~\eqref{eq:gfunction} as functions of the excess energy $Q$.
\label{fig:gfunction}}
\end{figure}

In Fig.~\ref{fig:gfunction}, we show the real and imaginary parts of the loop
function $G$ as a function of the excess energy $Q$ ($Q = \sqrt{s} - m_\eta -
M_{^3{\rm He}}$).

In the normalization that we are using, the $\eta$-nucleon and $\eta$-$^3$He
scattering lengths are related to the $t$-matrices by
\begin{eqnarray}
a_{\eta N} &=& \left.\frac{1}{4\pi}\frac{m_N}{\sqrt{s_{\eta N}}} t_{\eta N}
\right|_{\sqrt{s_{\eta N}} = m_N + m_\eta}\\
a_{\eta ^3{\rm He}} &=& \left.\frac{1}{4\pi}\frac{M_{^3{\rm He}}}{\sqrt{s}}
T \right|_{\sqrt{s} = M_{^3{\rm He}} + m_\eta}.    \label{eq:aetahe3}
\end{eqnarray}

The strategy that we adopt is to fit $\tilde{V}$ to the $pd (d p)
\to \eta ^3{\rm He}$ data and then see how different $\tilde{V}$ is
from $3t_{\eta N}$ by evaluating
\begin{eqnarray}
a'_{\eta N} = \left.\frac{1}{4\pi}\frac{m_N}{\sqrt{s_{\eta N}}}
\frac{\tilde{V}}{3} \right|_{\sqrt{s_{\eta N}} = m_N + m_\eta}
\label{effective}
\end{eqnarray}
and comparing it to the theoretical value of $a_{\eta N}$.

After obtaining the \emph{best} value for $\tilde{V}$, we then plot
\begin{eqnarray}
T = \tilde{T} e^{-2\beta^2 |\vec{p}_{\eta}|^2}
\end{eqnarray}
and investigate $|T|^2$ below threshold to identify a bump and its width.
From this we can determine roughly the position and width of the bound state.
The more precise determination is done by plotting ${\rm Re}(T)$ and ${\rm
Im}(T)$, and we see that in a narrow range of $Q$ they are consistent with a
Breit Wigner form
\begin{eqnarray}
\label{eq:bwform}
T &=& \frac{g^2}{\sqrt{s} - M_R + i \Gamma/2}\\ &=& \frac{g^2(\sqrt{s} -
M_R)}{(\sqrt{s} - M_R)^2 + \Gamma^2/4} - i \frac{g^2\Gamma/2}{(\sqrt{s} -
M_R)^2 + \Gamma^2/4}, \nonumber
\end{eqnarray}
where $g$ is constant and $M_R$ and $\Gamma$ are the mass and width
of the $\eta ^3{\rm He}$ bound state.

\section{Production amplitude in the $\boldsymbol{s}$-wave}

Following the formalism of Ref.~\cite{Colin}, we write for the $pd
\to \eta ^3{\rm He}$ transition depicted as a circle in
Fig.~\ref{fig:fig1}
\begin{eqnarray}
V_P = A \vec{\epsilon} \cdot \vec{p} + i B (\vec{\epsilon} \times
\vec{\sigma} ) \cdot \vec{p},      \label{eq:vp}
\end{eqnarray}
where $\vec{\epsilon}$ is the polarization of the deuteron,
$\vec{\sigma}$ denotes the Pauli matrix standing for the spin of the
proton, and $\vec{p}$ is the momentum in the initial state. This
amplitude has the initial-state $p$-wave needed to match the $\eta
(0^-) ^3{\rm He} (1/2^+)$ with the $d (1^+) p (1/2^+)$ system. In
Ref.~\cite{Colin} $A=B$ was taken, which is consistent with the
SPESIV experiment~\cite{Berger:1988ba}, but there is no loss of
generality for the total cross section if another choice is made.

Some extra information on these parameters is obtained from
Ref.~\cite{Papenbrock:2014hup}, where the ratio $|A|/|B|$ was found to be
constant of the order of $0.9$ for $Q<10$~MeV. The choice $A=B$ can be
interpreted as having the $dp$ system in spin $1/2$, according to the study
done in Ref.~\cite{Lu:2016roh}, where the analogous conjugate reaction
$\Lambda_b \to J/\psi K^0 \Lambda$ was studied.

With similar arguments to those used to derive Eq.~\eqref{eq:opotential}, we
can justify that $V_P$ in Eq.~\eqref{eq:vp} must be accompanied by the factor
$e^{-\beta^2|\vec{p}_\eta|^2}$, which, if the $\eta$ is in the loop, will
become $e^{-\beta^2|\vec{q}|^2}$. In view of this we can write analytically
the equation for the diagrams of Fig.~\ref{fig:fig1} as,
\begin{eqnarray}
t_{pd \to \eta ^3{\rm He}} &=& V_P e^{-\beta^2 |\vec{p}_{\eta}|^2} +
V_P G \tilde{T} e^{-\beta^2 |\vec{p}_{\eta}|^2}\nonumber\\ &=& V_P
e^{-\beta^2 |\vec{p}_{\eta}|^2} (1 + G \tilde{T}) =  \frac{V_P
e^{-\beta^2 |\vec{p}_{\eta}|^2}}{1-\tilde{V}G},
\label{eq:tdptoetahe3}
\end{eqnarray}
where in the last step we have used Eq.~\eqref{eq:BSE}. The cross section
then becomes
\begin{eqnarray}
\sigma = \frac{m_p M_{^3{\rm He}}}{12 \pi s} (|A'|^2 + 2|B'|^2)
|\vec{p}_\eta| |\vec{p}| e^{-2\beta^2 |\vec{p}_{\eta}|^2},
\label{eq:tcs-swava}
\end{eqnarray}
with
\begin{eqnarray}
A' = \frac{A}{1-\tilde{V}G}; ~~~~~~~~~~ B' = \frac{B}{1-\tilde{V}G}.
\end{eqnarray}
This means that, in the analysis of the total cross section with only
$s$-waves, we can, without loss of generality, take $A=B$ and real. This will
allow us to perform a fit to the data up to an excess energy $Q=2$~MeV, where
the $p$-wave effects are negligible, and thus determine $\tilde{V}$. From
these, by means of Eq.~\eqref{eq:ttilde}, we shall determine $T$ and
investigate its structure below threshold.

\section{The inclusion of $\boldsymbol{p}$-waves}

Following once again the approach of Ref.~\cite{Colin}, we assume an
$\eta$-$^3{\rm He}$ $p$-wave production amplitude
\begin{eqnarray}
V_{1P} = C \vec{\epsilon} \cdot \vec{p}_\eta + i D (\vec{\epsilon}
\times \vec{\sigma} ) \cdot \vec{p}_\eta .      \label{eq:v1p}
\end{eqnarray}
This amplitude will be taken empirically and once again there is support for
$C=D$ from the experiment of Ref.~\cite{Berger:1988ba}. Hence we take $A = B$
and $C = D$ as in Ref.~\cite{Colin}.

As in Ref.~\cite{Colin} we shall also take into account the $s$- and $p$-wave
interference by means of the asymmetry parameter $\alpha$ defined as
\begin{eqnarray}
\alpha = \frac{\dd}{\dd\cos\theta_\eta}\ln\left.\left(\frac{\dd\sigma}{\dd\Omega}\right)\right|_{\cos\theta_\eta = 0}.
\end{eqnarray}
By means of Eqs.~\eqref{eq:tdptoetahe3} and \eqref{eq:v1p} we obtain
\begin{eqnarray}
\frac{\dd\sigma}{\dd\Omega}  &= & \frac{m_p M_{^3{\rm He}}}{48 \pi^2 s}
\frac{|\vec{p}_\eta|}{|\vec{p}|} \left( (|A'|^2 + 2|B'|^2)
|\vec{p}|^2 e^{-\beta^2 |\vec{p}_{\eta}|^2} \right.\\  \nonumber
&& \hspace{-2cm}\left.\phantom{\int} + (|C|^2 + 2|D|^2)
|\vec{p}_\eta|^2+ 2 {\rm Re}(A'C^* + 2B'D^*) |\vec{p}||\vec{p}_\eta| \cos(\theta_{\eta}) \right),
\end{eqnarray}
from which we find that
\begin{eqnarray}
\alpha = \frac{2 {\rm Re}(A'C^* + 2B'D^*) |\vec{p}||\vec{p}_\eta|}{
(|A'|^2 + 2|B'|^2) |\vec{p}|^2 e^{-2\beta^2 |\vec{p}_{\eta}|^2} +
(|C|^2 + 2|D|^2) |\vec{p}_\eta|^2 }. \label{eq:asymmetry-alpha}
\end{eqnarray}

In addition, the total cross section of Eq.~\eqref{eq:tcs-swava} becomes
\begin{eqnarray}
\sigma &=& \frac{m_p M_{^3{\rm He}}}{12 \pi s}
\frac{|\vec{p}_\eta|}{|\vec{p}|} \left( (|A'|^2 + 2|B'|^2)
|\vec{p}|^2 e^{-2\beta^2 |\vec{p}_{\eta}|^2}\right.\nonumber\\ &+& \left.(|C|^2 + 2|D|^2)
|\vec{p}_\eta|^2 \right). \label{eq:tcs-spwava}
\end{eqnarray}
Equations~\eqref{eq:asymmetry-alpha} and \eqref{eq:tcs-spwava} are
used to fit the experimental data on the $dp \to \eta ^3{\rm He}$
total cross section and asymmetry parameter $\alpha$ of
Ref.~\cite{Mersmann}.

The value of $\tilde{V}$ obtained from the $s$-wave analysis was
used as a starting value for the global fit but the resulting
parameter does not differ significantly from that found in the
$s$-wave analysis.

\section{Results}

\subsection{$\boldsymbol{s}$-wave analysis}

First, we perform three-parameter ($A = B = r_A$ and $\tilde{V} =
{\rm Re}(V) + i {\rm Im}(V)$) $\chi^2$ fits to the experimental data
on the  total cross sections of the $pd (dp) \to \eta ^3{\rm He}$
reaction below $Q = 2$~MeV. The main purpose of this part of the fit
is to provide starting values for the parameters of the global fit,
where the full measured range in $Q$ is used.

\subsection{Results including the $\boldsymbol{p}$-wave}

We next perform six-parameter ($A = B = r_A$, $C=D = r_Ce^{i\theta}(1 +
\gamma Q)$, and $\tilde{V} = {\rm Re}(V) + i {\rm Im}(V)$) $\chi^2$ fits to
the experimental data on the  total cross sections and asymmetry of the $pd
\to \eta ^3{\rm He}$ reaction~\cite{Mersmann}. The values of the resulting
parameters are collected in Table~\ref{tab:fittedparas}.

\begin{table}[htbp]
\caption{Values of parameters determined in this
work.}\label{tab:fittedparas}
\begin{tabular}{|c|c|}
\hline
Parameter   & Fitted value \\
\hline
 $r_A [{\rm MeV}^{-2}]$   & $(9.43 \pm 0.17) \times 10^{-7}$ \\
 $r_C [{\rm MeV}^{-2}]$   & $(6.85 \pm 0.31) \times 10^{-6}$ \\
 $\theta [{\rm degree}]$   & $347 \pm 2$ \\
 $\gamma [{\rm MeV}^{-1}]$           & $(-5.25 \pm 0.15)\times 10^{-2}$     \\
 ${\rm Re}(V) [{\rm MeV}^{-1}]$      & $(-14.57 \pm 0.42)\times 10^{-2}$    \\
 ${\rm Im}(V) [{\rm MeV}^{-1}]$       & $(-5.36 \pm 0.14)\times 10^{-2}$    \\
\hline
\end{tabular}
\end{table}

With the potential $\tilde{V}$ obtained from these fits, we have evaluated
the scattering length $a'_{\eta N}$ of Eq.~\eqref{effective}:
\begin{eqnarray}
a'_{\eta N} = [-(0.48 \pm 0.05) - i(0.18 \pm 0.02) ] ~~ {\rm fm}.
\label{eq:aetaNresults}
\end{eqnarray}
These results are very close to those of the $a_{\eta N}$ scattering length
determined in different studies, and fair enough for the approximation
introduced in the low density theorem of Eq.~\eqref{eq:vroptical}, thus
giving support to the analysis done here.

Similarly, by means of Eq.~\eqref{eq:aetahe3}, we determine the $\eta ~
^3{\rm He}$ scattering length to be
\begin{eqnarray}
a_{\eta ^3{\rm He}} = [(2.23 \pm 1.29) - i(4.89 \pm 0.57) ] ~~ {\rm
fm}.   \label{eq:aetahe3results}
\end{eqnarray}

Note that the strategy of fitting an optical potential to the data
instead of the usual $t$-matrix parametrization used in previous
works, allows us to determine the sign of the real part of the
scattering lengths. The fit yields an attractive potential, which is
consistent with all theoretical derivations of $t_{\eta N}$,
together with the $t_{\eta N} \tilde{\rho}(\vec{r})$ assumption for
the optical potential.

It is interesting to see that the errors for $a_{\eta ^3{\rm He}}$
are relatively large. This is in fact not surprising, since in
Ref.~\cite{Colin} $a_{\eta ^3{\rm He}} = (\pm 10.9 - i 1.0) ~{\rm
fm}$ was obtained from the ANKE data, while in
Ref.~\cite{Smyrski:2007nu} the COSY-11 collaboration reported
$a_{\eta ^3{\rm He}} = (\pm 2.9 - i 3.2) ~{\rm fm}$, though without
taking beam smearing effects into account. Nevertheless, the two raw
data sets are compatible.

It is also interesting to mention that results for $a_{\eta ^3{\rm
He}}$ similar to those in Eq.~\eqref{eq:aetahe3results}, $(2.3,
-i3.2)$ fm, were found from the simultaneous analysis of much cruder
data for the $pd \to \eta ^3{\rm He}$ and $dd \to \eta ^4{\rm He}$
reactions, making fits in terms of an optical
potential~\cite{Willis:1997ix}.

\begin{figure}[htbp]\centering
\includegraphics[width=1.0\columnwidth]{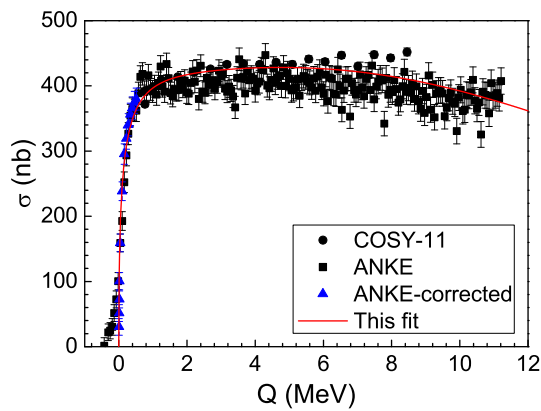}
\caption{The fitted $dp \to \eta ^3{\rm He}$ total cross sections
compared with experimental data~\cite{Mersmann,Smyrski:2007nu}.
\label{fig:tcsfull}}
\end{figure}

\begin{figure}[htbp]\centering
\includegraphics[width=1.0\columnwidth]{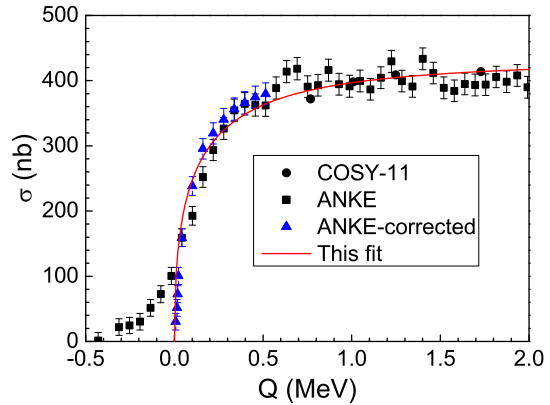}
\caption{The fitted $dp \to \eta ^3{\rm He}$ total cross sections
below $Q = 2$~MeV compared with experimental
data~\cite{Mersmann,Smyrski:2007nu}. \label{fig:tcsbelow2mev}}
\end{figure}

The fitted total cross sections reproduce well the experimental data shown in
Fig.~\ref{fig:tcsfull}. In Fig.~\ref{fig:tcsbelow2mev} we show the detail of
this fit in the $Q < 2$~MeV region. Note that Eq.~\eqref{eq:tcs-spwava} is
only valid for $Q>0$. Due to experimental resolution and beam momentum
spread, some data in Figs.~\ref{fig:tcsfull} and \ref{fig:tcsbelow2mev}
appear below threshold. These data are corrected by inverting the implicit
convolution of the real cross sections with the experimental resolution as
carried out by the ANKE collaboration (see more details in
Ref.~\cite{Mersmann}). The corrected data are shown in
Figs.~\ref{fig:tcsfull} and \ref{fig:tcsbelow2mev} by blue triangles. These
are obtained by shifting the experimental data to the deconvoluted
distribution of Ref.~\cite{Mersmann}.

\begin{figure}[htbp]\centering
\includegraphics[width=1.0\columnwidth]{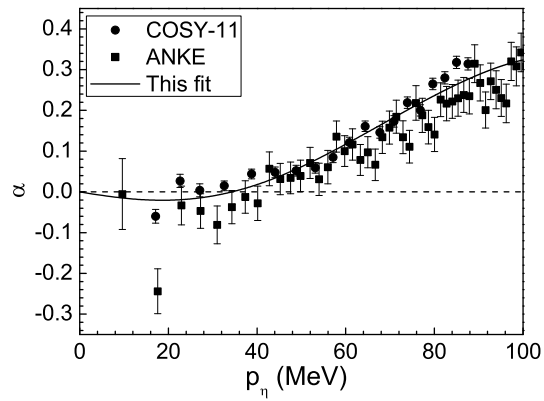}
\caption{The fit in the model to the asymmetry parameter $\alpha$ as a
function of the center-of-mass $\eta$ momentum $p_{\eta}$ compared with the
experimental data~\cite{Mersmann,Smyrski:2007nu}. \label{fig:asymmetry}}
\end{figure}

The fitted results of the asymmetry parameter $\alpha$ shown in
Fig.~\ref{fig:asymmetry} also describe well the experimental data.

\begin{figure}[htbp]\centering
\includegraphics[width=1.0\columnwidth]{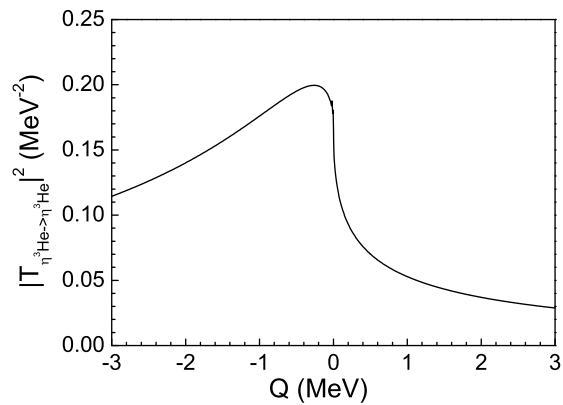}
\caption{Square of the absolute value of the $\eta ^3{\rm He} \to \eta ^3{\rm
He}$ scattering amplitude. \label{fig:tsquare}}
\end{figure}

We next turn our attention to the $\eta ^3{\rm He} \to \eta ^3{\rm He}$
scattering amplitude. In Fig.~\ref{fig:tsquare}, we depict $|T|^2$ obtained
with the fitted parameters given in Table~\ref{tab:fittedparas}, as a
function of $Q$. We see a qualitative picture of a very weakly bound state of
$\eta ^3{\rm He}$ system.

\begin{figure}[htbp]\centering
\includegraphics[width=1.0\columnwidth]{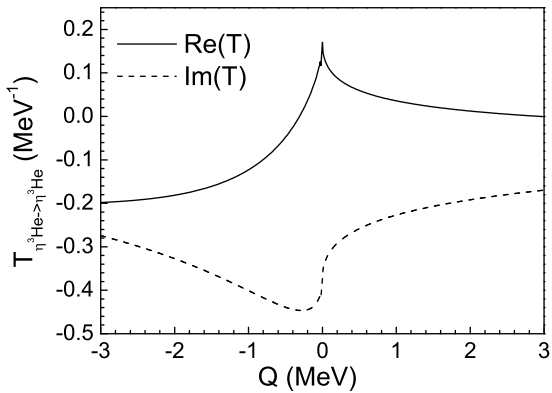}
\caption{Real and imaginary parts of the $\eta ^3{\rm He} \to \eta ^3{\rm
He}$ amplitude $T$ as a function of the excess energy $Q$.
\label{fig:tetahe3}}
\end{figure}

In order to make more quantitative statements, we plot in
Fig.~\ref{fig:tetahe3} the real and imaginary parts of $T$. We see
that at $Q = -0.3$~MeV ${\rm Re}(T)$ goes from negative to positive
passing through zero, ${\rm Im}(T)$ is negative, and $-{\rm Im}(T)$
has a peak. In a narrow window around $0.3$~MeV this amplitude has
the simple Breit-Wigner form of Eq.~\eqref{eq:bwform} with an energy
$M_R$ corresponding to a binding $B_E = 0.3 \pm 0.1$~MeV and a width
$\Gamma = 3.0 \pm 0.5$~MeV. The errors quoted here are derived from
those of the fitted potential listed in Table~\ref{tab:fittedparas}.

\begin{figure}[htbp]\centering
\includegraphics[width=1.0\columnwidth]{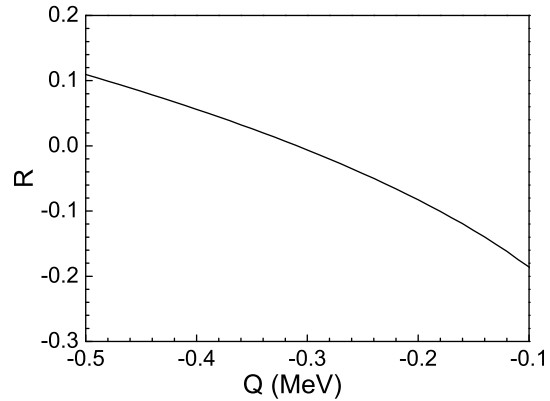}
\caption{Ratio of the real and imaginary parts of the $\eta ^3{\rm He} \to
\eta ^3{\rm He}$ amplitude $T$ as a function of the excess energy $Q$.
\label{fig:ratio}}
\end{figure}

The binding is determined from the zero of ${\rm Re}(T)$ and the width from
${\rm Im}(T)$. For this we write the ratio of the real and imaginary parts of
$T$ which, following Eq.~\eqref{eq:bwform}, become:
\begin{eqnarray}
R = \frac{{\rm Re}(T)}{{\rm Im}(T)} = -\frac{2}{\Gamma} (Q + B_E),
\end{eqnarray}
where $B_E$ is the binding energy (positive) of the $\eta ^3{\rm
He}$ system. Our results for $R$ shown in Fig.~\ref{fig:ratio} allow
us to estimate easily the values of $B_E$ and $\Gamma$. The binding
is very small and the width is much larger than the binding, in line
with other theoretical studies of $\eta$ bound in nuclei.

\subsection{Systematic uncertainties}

Two tests have been carried out to provide some indication of the systematic
uncertainties and the stability of the results.

Although the nuclear density used to determine the $G$ function of
Eq.~\eqref{eq:gfunction} is one of the important points of the analysis, we
have tried a $G$ function with a cut off of $300 - 400$~MeV and conducted the
fit again. It is well known that changes in $G$ can be approximately absorbed
in $\tilde{V}$ to obtain the same $T$ matrix. This is the case here and we
find that such large changes in $G$ induce only changes of the order of
$0.05$~MeV in the binding energy and $0.4$~MeV in the width.

Another test that we have undertaken is to assume an energy dependence of the
parameters $A$ and $B$ that could arise from the basic in $pd \to \eta ^3{\rm
He}$ production mechanism. For this we have taken a form $(1+\gamma' Q)$ with
$\gamma'$ sufficiently small that for $Q < 10$~MeV the changes in $|A|^2$ and
$|B|^2$ did not exceed 20\%. The resulting changes in the binding energy and
width were very small, being of the same order of magnitude as those induced
by the changes in $G$. Compounding these systematic effects in quadrature, we
conclude that
\begin{eqnarray}
B_E &=& (0.30 \pm 0.10 \pm 0.08)~{\rm MeV}, \label{eq:binding} \\
\Gamma &=& (3.0 \pm 0.5 \pm 0.7)~{\rm MeV}, \label{eq:Gamma-theory}
\end{eqnarray}
where the first errors are statistical and the second systematic.

\section{Considerations on poles of the amplitude}

Let us suppose that a microscopic theory, where the decay channels are known,
produces an amplitude like that in Eq.~\eqref{eq:bwform}, which corresponds
to a resonance. It is customary to assume that this amplitude has a pole at
$\sqrt{s} = M_{R} - i \Gamma/2$, and this would be true if $\Gamma$ were
constant. In a dynamical theory this no longer follows. To see this, consider
the case of an $s$-wave resonance where there is just one decay channel and
$\Gamma$ is proportional to the momentum $p$ of the decay products in the
resonance rest frame. If $\sqrt{s}$ is made complex, then so is $p$ and one
does not find a solution to $\sqrt{s} - M_{R} + i \Gamma/2 = 0$. A solution
may be obtained by changing $p \to - p$, which defines the second Riemann
sheet.

However, when one has only an optical potential, as in the present case, one
does not know the explicit channels contributing to the imaginary part and
their strength and the different Riemann sheets are not defined. The only
possibility to find poles is to change $\sqrt{s}$ to a complex value, with
the same optical potential, and look for a pole of the amplitude.

Imagine that we have an amplitude of the Breit-Wigner type of
Eq.~\eqref{eq:bwform}, valid in a range of values of $\sqrt{s}$, let us say
$M_R \pm \Gamma$. In this case the amplitude has a pole at $\sqrt{s} = M_{R}
- i \Gamma/2$. Since our amplitude behaves like a Breit Wigner in a certain
range of energies (see Fig.~\ref{fig:tetahe3}), we might think that it has a
pole at $-B_E - i \Gamma/2$ (see Eqs.~\eqref{eq:binding} and
\eqref{eq:Gamma-theory}). However, for this to be true the range of the
validity of the formula should stretch in an interval of about $6$ MeV, when
in reality it is barely valid in a range of $0.3$ MeV. As a consequence, we
do not find a pole at $-B_E - i \Gamma/2$ and, instead, we find it at $Q =
(1.5-i0.7)$~MeV, i.e., in the unbound region.

With the definition that a bound state corresponds to a pole of the amplitude
below threshold, the potential that we obtain would not produce a bound
state. Solving the Schr\"odinger equation with bound-state boundary
conditions would not have led to a solution since it would be equivalent to
having a pole below threshold. This is indeed the situation of
Ref.~\cite{Barnea:2015lia} where, for values of ${\rm Re}(a'_{\eta N}) \simeq
-0.5$ fm, as we found in Eq.~\eqref{eq:aetaNresults}, they would not obtain a
bound state solution. We have checked that, by taking ${\rm Im}(a'_{\eta N})
\simeq -0.25$~fm we need ${\rm Re}(a'_{\eta N}) \simeq -0.66$~fm in order to
have a pole in the bound region at $Q = (-0.14 - i4.1)$~MeV. The model of
Ref.~\cite{Barnea:2015lia} requires an even larger ${\rm Re}(a'_{\eta N})$,
of the order of $1$~fm, to get bound states, but since they have a larger
imaginary part of the potential, the resulting widths are of the order of
$15$~MeV. Given that the imaginary part of a potential acts rather like a
repulsion, one can see consistency between their results and ours.

The potential that we have found does not produce a pole below threshold and
so, with the conventional definition, there is no bound state. The search
could stop at this point but this would be premature since one then misses
all the information that we have found about the amplitude below threshold,
as shown in Figs.~\ref{fig:tsquare} and \ref{fig:tetahe3}. If an
experimentalist observed a structure of $|T|^2$ similar to that in
Fig.~\ref{fig:tsquare} he would conclude that there was a bound state below
threshold. From the practical point of view this may have more relevance than
the existence of a pole in the unbound region and on a complex plane that is
not experimentally accessible. The fact that, with the potential that we have
derived, there is a pole in the unbound region, might lead one to think that
there is a resonance around that pole position. This would be the wrong
conclusion. We consider that it is more appropriate to study the peak
structure of the amplitude below threshold and this is the attitude we have
taken here.

\section{Summary and conclusions}

We have performed an analysis of data on the $pd (dp) \to \eta
^3{\rm He}$ reaction close to threshold. These consist of total
cross sections and angular asymmetries up to an excess energy of
$10$~MeV. Unlike former approaches that make a parametrization of
the amplitude, we express the observables in terms of an optical
potential from which the $\eta ^3 {\rm He}$ scattering amplitude is
evaluated. The $T$ matrix is evaluated from the potential using the
Bethe-Salpeter equation and the loop function $G$ of the
intermediate $\eta ^3 {\rm He}$ state. This reflects the range of
the $\eta ^3 {\rm He}$ interaction, as given by the empirical
density of the $^3 {\rm He}$ nucleus. The results lead to a
structure of the $T$ matrix that is quite different from the usual
parametrizations of the data.

The potential and other parameters related to the production
vertices are fitted to the data and in this way we deduce that there
is a weakly bound $\eta ^3{\rm He}$ state with binding energy of the
order of $0.3$~MeV and a width of the order of $3$~MeV. We also
obtain an $\eta ^3 {\rm He}$ scattering length of the order of
$(2.2-i4.9)~ {\rm fm}$. It is important to note that the fit in
terms of the potential resolves an ambiguity in the sign of the real
part of $a_{\eta ^3 {\rm He}}$ that was present in previous
analyses.

In summary, the new approach to the analysis of the $pd (dp) \to
\eta ^3 {\rm He}$ data close to threshold has proved quite useful
and has been able to provide information on a bound state of the
$\eta ^3 {\rm He}$ system. This agrees with previous theoretical
work that the width of the bound state should be significantly
larger than the binding. A peak in $|T|^2$ is predicted below
threshold and this might in principle be detected experimentally in
some breakup channel of the $pd$ system. However, one must
anticipate that the background from other sources, where the bound
state is not produced, could be very large and obscure any signal.

\section*{Acknowledgments}

Correspondence with A.~Gal has proved most useful in the writing of
this paper. One of us, E.~O., wishes to acknowledge support from the
Chinese Academy of Science in the Program of ``CAS President's
International Fellowship for Visiting Scientists'' (Grant No.\
2013T2J0012). This work is partly supported by the National Natural
Science Foundation of China (Grants No.\ 11565007, 11547307, and
No.\ 11475227) and the Youth Innovation Promotion Association CAS
(No.\ 2016367). This work is also partly supported by the Spanish
Ministerio de Economia y Competitividad and European FEDER funds
under the contract number FIS2011-28853-C02-01, FIS2011-
28853-C02-02, FIS2014-57026-REDT, FIS2014-51948-C2- 1-P, and
FIS2014-51948-C2-2-P, and the Generalitat Valenciana in the program
Prometeo II-2014/068. We acknowledge the support of the European
Community-Research Infrastructure Integrating Activity Study of
Strongly Interacting Matter (acronym HadronPhysics3, Grant Agreement
n. 283286) under the Seventh Framework Programme of the EU. M.~S.\
and P.~M.\ acknowledge support from the Polish National Science
Center through grants Nos.\ DEC-2013/11/N/ST2/04152 and
2011/01/B/ST2/00431.

\bibliographystyle{plain}

\begin{thebibliography}{999}

\bibitem{Wilkin:2016ajz}
  C.~Wilkin,
  Acta Phys.\ Polon.\ B {\bf 47}, 249 (2016).

\bibitem{Bass:2015ova}
  S.~D.~Bass and P.~Moskal,
  Acta Phys.\ Polon.\ B {\bf 47}, 373 (2016).

\bibitem{Haider:2015fea}
  Q.~Haider and L.~C.~Liu,
  Int.\ J.\ Mod.\ Phys.\ E {\bf 24}, 1530009 (2015).

\bibitem{Kelkar:2015bta}
  N.~G.~Kelkar,
  Acta Phys.\ Polon.\ B {\bf 46}, 113 (2015).

\bibitem{Hirenzaki:2015eoa}
  S.~Hirenzaki, H.~Nagahiro, N.~Ikeno, and J.~Yamagata-Sekihara,
  Acta Phys.\ Polon.\ B {\bf 46}, 121 (2015).

\bibitem{Friedman:2013zfa}
  E.~Friedman, A.~Gal, and J.~Mare${\rm \check{s}}$,
  Phys.\ Lett.\ B {\bf 725}, 334 (2013).

\bibitem{Bhaleliu}
  R.~S.~Bhalerao and L.~C.~Liu,
  Phys.\ Rev.\ Lett.\  {\bf 54}, 865 (1985).

\bibitem{Haiderliu}
  Q.~Haider and L.~C.~Liu,
  Phys.\ Lett.\ B {\bf 172}, 257 (1986).

\bibitem{Liuhaider}
  L.~C.~Liu and Q.~Haider,
  Phys.\ Rev.\ C {\bf 34}, 1845 (1986).

\bibitem{chiang}
  H.~C.~Chiang, E.~Oset, and L.~C.~Liu,
  Phys.\ Rev.\ C {\bf 44}, 738 (1991).

\bibitem{Kaiser}
  T.~Waas, N.~Kaiser, and W.~Weise,
  Phys.\ Lett.\ B {\bf 379}, 34 (1996).

\bibitem{ramos}
  E.~Oset and A.~Ramos,
  Nucl.\ Phys.\ A {\bf 635}, 99 (1998).

\bibitem{ollerulf}
  J.~A.~Oller and U.-G.~Mei{\ss}ner,
  Phys.\ Lett.\ B {\bf 500}, 263 (2001).

\bibitem{Nieves}
  C.~Garcia-Recio, J.~Nieves, E.~Ruiz Arriola, and M.~J.~Vicente Vacas,
  Phys.\ Rev.\ D {\bf 67}, 076009 (2003).

\bibitem{Hyodo}
  T.~Hyodo, S.~I.~Nam, D.~Jido, and A.~Hosaka,
  Phys.\ Rev.\ C {\bf 68}, 018201 (2003).

\bibitem{Inoue}
  T.~Inoue, E.~Oset, and M.~J.~Vicente Vacas,
  Phys.\ Rev.\ C {\bf 65}, 035204 (2002).

\bibitem{Jido}
  T.~Hyodo, D.~Jido, and A.~Hosaka,
  Phys.\ Rev.\ C {\bf 78}, 025203 (2008).

\bibitem{Sekihara}
  T.~Sekihara, T.~Hyodo, and D.~Jido,
  Prog.\ Theor.\ Exp.\ Phys.\ {\bf 2015}, 063D04 (2015).

\bibitem{Garzon}
  E.~J.~Garzon and E.~Oset,
  Phys.\ Rev.\ C {\bf 91}, 025201 (2015).

\bibitem{Inoueta}
  T.~Inoue and E.~Oset,
  Nucl.\ Phys.\ A {\bf 710}, 354 (2002).


\bibitem{GarciaRecio:2002cu}
  C.~Garcia-Recio, J.~Nieves, T.~Inoue, and E.~Oset,
  Phys.\ Lett.\ B {\bf 550}, 47 (2002).

\bibitem{Cieply:2013sga}
  A.~Ciepl\'{y}, E.~Friedman, A.~Gal, and J.~Mare\v{s},
  Nucl.\ Phys.\ A {\bf 925}, 126 (2014).

\bibitem{Barnea:2015lia}
  N.~Barnea, E.~Friedman, and A.~Gal,
  Phys.\ Lett.\ B {\bf 747}, 345 (2015).

\bibitem{Bilger:2002aw}
  R.~Bilger {\it et al.},
  Phys.\ Rev.\ C {\bf 65}, 044608 (2002).

\bibitem{Mersmann}
  T.~Mersmann {\it et al.},
  Phys.\ Rev.\ Lett.\  {\bf 98}, 242301 (2007).

\bibitem{Colin}
  C.~Wilkin {\it et al.},
  Phys.\ Lett.\ B {\bf 654}, 92 (2007).

\bibitem{Urban:2009zzc}
  J.~Urban {\it et al.} [GEM Collaboration],
  Int.\ J.\ Mod.\ Phys.\ A {\bf 24}, 206 (2009).

\bibitem{Chrien:1988gn}
  R.~E.~Chrien {\it et al.},
  Phys.\ Rev.\ Lett.\  {\bf 60}, 2595 (1988).

\bibitem{Johnson:1993zy}
  J.~D.~Johnson {\it et al.},
  Phys.\ Rev.\ C {\bf 47}, 2571 (1993).

\bibitem{Willis:1997ix}
  N.~Willis {\it et al.},
  Phys.\ Lett.\ B {\bf 406}, 14 (1997).

\bibitem{Sokol:1998ua}
  G.~A.~Sokol, T.~A.~Aibergenov, A.~V.~Kravtsov, A.~I.~L'vov, and L.~N.~Pavlyuchenko,
  Fizika B {\bf 8}, 85 (1999).

\bibitem{Budzanowski:2008fr}
  A.~Budzanowski {\it et al.} [COSY-GEM Collaboration],
  Phys.\ Rev.\ C {\bf 79}, 012201 (2009).

\bibitem{Moskal:2010ee}
  P.~Moskal and J.~Smyrski,
  Acta Phys.\ Polon.\ B {\bf 41}, 2281 (2010).

\bibitem{Pheron:2012aj}
  F.~Pheron {\it et al.},
  Phys.\ Lett.\ B {\bf 709}, 21 (2012).

\bibitem{Adlarson:2013xg}
  P.~Adlarson {\it et al.} [WASA-at-COSY Collaboration],
  Phys.\ Rev.\ C {\bf 87}, 035204 (2013).

\bibitem{Fujioka:2015pla}
  H.~Fujioka {\it et al.} [Super-FRS Collaboration],
  Acta Phys.\ Polon.\ B {\bf 46}, 127 (2015).

\bibitem{Skurzok:2016fuv}
  M.~Skurzok {\it et al.} [WASA-at-COSY Collaboration],
  Acta Phys.\ Polon.\ B {\bf 47}, 503 (2016).

\bibitem{Rausmann:2009dn}
  T.~Rausmann {\it et al.},
  Phys.\ Rev.\ C {\bf 80}, 017001 (2009).

\bibitem{Smyrski:2007nu}
  J.~Smyrski {\it et al.},
  Phys.\ Lett.\ B {\bf 649}, 258 (2007).

\bibitem{Wycech}
  A.~M.~Green and S.~Wycech,
  Phys.\ Rev.\ C {\bf 60}, 035208 (1999)

\bibitem{Batinic:1995yu}
  M.~Batini\'{c}, I.~\v{S}laus, and A.~\v{S}varc,
  Phys.\ Rev.\ C {\bf 52}, 2188 (1995).

\bibitem{Batinic:1995kr}
  M.~Batini\'c, I.~\v{S}laus, A.~\v{S}varc, and B.~M.~K.~Nefkens,
  Phys.\ Rev.\ C {\bf 51}, 2310 (1995),
  Erratum: Phys.\ Rev.\ C {\bf 57}, 1004 (1998).

\bibitem{Wycech:2014wua}
  S.~Wycech and W.~Krzemie${\rm \acute{n}}$,
  Acta Phys.\ Polon.\ B {\bf 45}, 745 (2014).

\bibitem{Sick:2014yha}
  I.~Sick,
  Phys.\ Rev.\ C {\bf 90}, 064002 (2014).

\bibitem{Berger:1988ba}
  J.~Berger {\it et al.},
  Phys.\ Rev.\ Lett.\  {\bf 61}, 919 (1988).

\bibitem{Papenbrock:2014hup}
  M.~Papenbrock {\it et al.},
  Phys.\ Lett.\ B {\bf 734}, 333 (2014).

\bibitem{Lu:2016roh}
  J.~X.~Lu, E.~Wang, J.~J.~Xie, L.~S.~Geng, and E.~Oset,
  Phys.\ Rev.\ D {\bf 93}, 094009 (2016).

\end{thebibliography}

\end{document}